\documentclass[
 reprint,
 amsmath,amssymb,
 aps,
 prl,
]{revtex4-1}

\usepackage{graphicx}
\usepackage{dcolumn}
\usepackage{bm}
\usepackage{subfig}
\usepackage{xcolor}
\usepackage{nicefrac}
\usepackage{chemfig}
\usepackage{tikz}
\usetikzlibrary{matrix,decorations.pathreplacing}
\newcommand{\mat}[1]{\tilde{#1}}
\newcommand{\op}[1]{\hat{#1}}
\newcommand{\change}[1]{\textcolor{black}{#1}}
\begin{document}


\title{Symmetry analysis of odd- and even-frequency superconducting gap symmetries for time-reversal symmetric interactions}
\author{R. Matthias Geilhufe$^1$}
\email{Geilhufe@kth.se}
\author{Alexander V. Balatsky$^{1,2,3}$}
\affiliation{$^1$Nordita, KTH Royal Institute of Technology and Stockholm University, Roslagstullsbacken 23, SE-106 91 Stockholm, Sweden\\
$^2$Institute for Materials Science, Los Alamos National Laboratory, Los Alamos, NM 87545, USA\\
$^3$Department of Physics, University of Connecticut, Storrs, CT 06269, USA}

\date{\today}

\begin{abstract}
Odd-frequency superconductivity describes a class of superconducting states where the superconducting gap is an odd function in relative time and Matsubara frequency. We present a group theoretical analysis based
on the linearized gap equation in terms of Shubnikov groups of the second kind. By discussing systems with spin-orbit coupling and an interaction kernel which is symmetric under the reversal of \change{relative time}, we show
that both even- and odd-frequency gaps are allowed to occur. Specific examples are discussed for the square lattice, the octahedral lattice, and the tetragonal lattice. For irreducible representations that are even
under \change{reversal of relative time} the common combinations of $s$- and $d$-wave spin singlet and $p$-wave spin triplet gaps are revealed, irreducible representations that are odd under \change{reversal of relative time} give rise to $s$- and $d$-wave spin triplet and $p$-wave spin singlet gaps. Furthermore, we discuss the construction of a generalized Ginzburg-Landau theory in terms of the associated irreducible representations. The result complements the established classification of superconducting states of matter.
\end{abstract}

\maketitle
Unconventional superconductors such as the heavy fermion systems, e.g., CeCu$_2$Si$_2$ \cite{pfleiderer2009superconducting,steglich1979superconductivity}, Sr$_2$RuO$_{4}$ \cite{nelson2004odd}, and UPt$_3$ \cite{strand2010transition,joynt2002superconducting}; the cuprates, e.g., YBa$_2$Cu$_3$O$_7$ \cite{PhysRevLett.58.908} and HgBa$_2$CuO$_{4+\delta}$ \cite{schilling1993superconductivity}; and the organic superconductors like the BEDT-TTF-based charge transfer salts \cite{commeau2017,Tajima2002SC,kamarchuk1990intramolecular,kobayashi1987crystal,urayama1988new} exhibit symmetries of the superconducting gap beyond the conventional BCS $s$-wave \cite{bardeen1957theory}. In this connection, a group theory analysis based on the underlying symmetry of the pairing potential is crucial in establishing an unified classification of the arising superconducting states of matter \cite{volovik1984unusual,volovik1985superconducting,blount1985symmetry,ueda1985p,sigrist1991phenomenological,sigrist1987symmetry}. In general, the pairing wave function of two electrons has to be anti-symmetric under particle interchange leading to the Pauli-principle. At equal times, and by neglecting orbital degrees of freedom two cases can occur: first, a gap odd in spin and even in parity, such as spin singlet $s$- and $d$-wave gaps, and, second, a gap even in spin and odd in parity, such as spin triplet $p$- and $f$-wave gaps.

However, as pointed out by Berezinskii \cite{berezinskii1974new} and Balatsky and Abrahams \cite{balatsky1992new} a pairing of particles beyond the conventional ones is possible, if the particle-particle correlator is zero at equal times but non-zero otherwise. This is achieved when the superconducting gap is an odd function in \change{relative time}, leading to the notion of odd-time or odd-frequency superconductivity, respectively \change{(odd-frequency also refers to gap functions odd in Matsubara frequency)}. Among others, odd-frequency contributions were reported to occur in connection to diffusive ferromagnet/superconductor junctions \cite{yokoyama2007manifestation,Volkov2003}, normal-metal/superconductor junctions \cite{tanaka2007odd}, topological insulators \cite{black2012odd}, heterostructures of transition-metal dichalcogenides and $s$-wave superconductors \cite{triola2016a}, multi-band superconductors \cite{black2013odd}, and driven systems \cite{Triola2016,triola2017pair}. Also, odd-frequency states were discussed in connection to time-reversal topological superconductivity in double Rashba wires, where it was found that odd-frequency pairing is strongly enhanced in the topological state \cite{Ebisu2016}. For some of the above mentioned systems, the respective signatures of odd-frequency correlations could also be verified experimentally \cite{di2015signature,PhysRevX.5.041021,pal2017spectroscopic, PhysRevLett.110.107005}. An extensive discussion of edge-states and topology in superconductors including odd-frequency gap functions was communicated by Tanaka, Sato and Nagaosa \cite{tanaka2011symmetry}. A comprehensive review on odd-frequency superconductivity with the overview of possible realizations is given in Ref. \cite{linder2017odd}.

In general, odd-frequency superconductivity can only occur when retardation is explicitly taken into account. Close to the superconducting transition temperature, the underlying gap equations can be linearized, leading to the so-called linearized gap equation or Bethe-Salpeter equation \cite{riseborough2004heavy}. By solving for the eigenvalues of the Bethe-Salpeter equation, odd-frequency solutions were found numerically, for example, in the repulsive Hubbard model \cite{Bulut1993,PhysRevB.79.174507} and also in organic charge-transfer salts \cite{PhysRevB.59.R713}. In such models with strong on-site repulsion, pairs can avoid the repulsion either by exhibiting a pair wave function with zero on-site amplitude, i.e., with nonzero angular momentum, or by establishing an odd-$\omega$ dependence which implies a vanishing equal-time pair amplitude.

Here, we extend the formalism of Refs. \cite{volovik1984unusual,volovik1985superconducting,blount1985symmetry,ueda1985p,sigrist1991phenomenological,sigrist1987symmetry} and show how a symmetry analysis of the Bethe-Salpeter equation can be performed by explicitly incorporating \change{reversal of relative time similarly to the construction of~}Shubnikov groups of the second kind. Since a solution of the Bethe-Salpeter equation, i.e., a superconducting gap function, transforms as one of the irreducible representations of the underlying symmetry group, it generally breaks certain symmetries of the pairing potential.  Therefore, gap functions that are odd under time-reversal symmetry can naturally occur even if the pairing potential itself is time-reversal symmetric. 

The paper is structured as follows. First, we introduce the formalism by introducing the Bethe-Salpeter equation and the transformation behavior of the superconducting gap. Then, we summarize the construction of Shubnikov groups, together with the construction of faithful representations needed for the calculation of their character tables. Afterwards, we show possible superconducting gap symmetries for the examples of the square lattice, the cubic lattice, as well as the non-centrosymmetric tetragonal lattice. In the last part, we discuss the generalized Ginzburg-Landau theory for even- and odd-frequency superconductors.

\section{Linearized gap equation and transformation behavior of the superconducting gap}
For a standard BCS approach, the superconducting gap is taken as frequency independent. Since the anomalous Green's function vanishes at zero time for odd-frequency superconductivity \cite{balatsky1994properties,abrahams1995properties} we have to choose a formalism incorporating a summation over time or frequency, respectively. Therefore, we stick to the Eliashberg formalism which is valid in the the strong-coupling regime \cite{PhysRev.148.263}. In general, the underlying Eliashberg equations which need to be solved self-consistently are nonlinear. However, in the region close to the superconducting transition temperature $T\approx T_c$, a corresponding linear equation can be formulated, which is called the linearized Eliashberg or Bethe-Salpeter equation. In the most general form it can be written as follows \cite{riseborough2004heavy,PhysRevB.79.174507,PhysRevB.59.R713},
\begin{multline}
 v \Delta_{\alpha\beta}(\vec{k},i\omega_n) = - \frac{T}{N} \sum_{\gamma,\delta} \sum_{\vec{k}'} \sum_m \Gamma_{\alpha\beta\gamma\delta}(\vec{k},\vec{k}',i\omega_m,i\omega_n)\\ \times G_\gamma(\vec{k}',i\omega_m)G_\delta(-\vec{k}',-i\omega_m) \Delta_{\gamma\delta}(\vec{k}',i\omega_m).
 \label{gap_equation}
\end{multline}
Here, $\Gamma_{\alpha\beta\gamma\delta}$ denotes the interaction kernel, the specific form of which depends on the system under consideration, e.g., electron-phonon interaction or a Berk-Schriefer-like interaction mediated by spin fluctuations \cite{monthoux1994self,PhysRevLett.17.433}, to name but a few. Furthermore, Greek indices denote the spin components, $G_\alpha$ is the Green's function for a particle with spin $\alpha$, $\Delta_{\alpha,\beta}$ is the superconducting gap, and $N$ denotes the total number of momenta in the Brillouin zone. The eigenvalue $\nu$ corresponds to a generalization of the linearized Eliashberg formalism allowing for multiple solutions of Eq. \eqref{gap_equation}, where a physical interpretation is only valid when the largest eigenvalue equals $\nu=1$, indicating a superconducting transition exhibiting the respective superconducting gap corresponding to $\nu$. Yet, the knowledge of the competing eigenvalues even if not physically realized in the system of interest gives an important insight into the allowed superconducting instabilities. Additionally, even in the nonlinear regime the symmetry of the solutions of the linearized equation can be used to study admixed phases as described in Ref. \cite{wojtanowski1986admixture,Leggett,sigrist1991phenomenological}. As Eq. \eqref{gap_equation} is a linear eigenvalue equation, it can be written as $v \Delta = \op{V}\,\Delta$, where $\op{V}$ denotes the kernel
\begin{multline}
V_{\alpha\beta\gamma\delta}(\vec{k},\vec{k}',i\omega_m,i\omega_n) = \Gamma_{\alpha\beta\gamma\delta}(\vec{k},\vec{k}',i\omega_m,i\omega_n) \\ \times G_\gamma(\vec{k}',i\omega_m)G_\delta(-\vec{k}',-i\omega_m).
\end{multline}
It is assumed that the symmetry of the crystal is reflected in the kernel $V$ and described by the symmetry group $\mathcal{G}$. Each eigenvector of Eq. \eqref{gap_equation} transforms as a basis function of an irreducible representation $\Gamma^p$ of $\mathcal{G}$ and the degeneracy of the corresponding eigenvalue is determined by the dimension of $\Gamma^p$, which will be denoted by $d_p$. Hence, the linearized gap equation can be reformulated as  
\begin{equation}
 v^{p,\nu} \mat{\Delta}_{m}^{p,\nu} = \op{V} \tilde{\Delta}_{m}^{p,\nu},
 \label{gap_equation2}
\end{equation}
where $m=1,\dots,d_p$ and $\nu=1,2,\dots$ counts over the multiple nonequivalent subspaces transforming as the same irreducible representation. A superconducting instability with a gap transforming as an irreducible representation $\Gamma^p$ occurs, if the corresponding eigenvalue $v^{p,\nu}$ reaches 1. Even though the pairing potential is invariant under every symmetry transformation of the group $\mathcal{G}$, the dominating gap function itself is only invariant under a subgroup, represented by one of the irreducible representations of $\mathcal{G}$.

It is assumed that the gap function transforms similarly to a pairing wave function. Considering spin-orbit coupling, each rotation in space (proper or improper) is connected to a specific rotation in spin space. Due to spin-orbit coupling, the single-particle states cannot be eigenstates of the spin operator in general, but can be labeled as pseudo-spin-states in a similar manner \cite{sigrist1991phenomenological}. The pseudo-spin-state is generated from a spin eigenstate by turning on the spin-orbit interaction adiabatically, leading to a one-to-one correspondence between the
original spin state and the pseudo-spin-state. Here we discuss the situation of having two states ($\uparrow$, $\downarrow$) similarly to the ordinary spin, where the transformations in the pseudo-spin space (or just spin space in the following) are generated by the Pauli matrices.

Applying the transformation operator $\op{g}$ associated to a specific symmetry transformation $g\in\mathcal{G}$ gives
\begin{equation}
 \op{g} \mat{\Delta}(\vec{k}) = \mat{u}^T(g) \mat{\Delta}\left(\mat{R}^{-1}(g)\vec{k}\right) \mat{u}(g).
  \label{rot_gap}
\end{equation}
Here, $\mat{R}(g)\in O(3)$ denotes a three-dimensional rotation matrix and $\mat{u}(g)\in SU(2)$ denotes the corresponding rotation matrix in spin space.
The concept of odd- and even-frequency gaps relates to the gap function being an odd- or even function of relative time or Matsubara frequency. 
\change{For a given gap function $\mat{\Delta}(\vec{k},t_1,t_2)$, the relative time-reversal operator $\op{T}$ acts by permuting the times $t_1$ and $t_2$. As shown in Appendix B, a similar representation of $\op{T}$ is found by applying a combination of matrix transpose, and $\vec{k}\rightarrow-\vec{k}$,
\begin{equation}
 \op{T} \mat{\Delta}(\vec{k}) = -\mat{\Delta}^T\left(-\vec{k}\right).
 \label{tr_gap}
\end{equation}}
Doing so, the action of $\op{T}$ can be discussed without explicitly taking into account \change{$t_1$ and $t_2$.}

With respect to the interchange of the spin indices within the gap function, mediated by the operator $\op{S}$, the gap function can be considered to be odd (singlet) or even (triplet). The resulting form of the gap in these cases is given by the antisymmetric matrix
\begin{equation}
 \mat{\Delta}(\vec{k}) = i \Psi(\vec{k}) \mat{\sigma}^y,
\end{equation}
for the spin singlet and by the symmetric matrix
\begin{equation}
 \mat{\Delta}(\vec{k}) = i \left(\vec{d}(\vec{k})\cdot\vec{\sigma}\right) \mat{\sigma}^y,
\end{equation}
for the spin triplet. Following Eqs. \eqref{rot_gap} and \eqref{tr_gap}, the transformation under group elements $\op{g}$ and under \change{relative time reversal} $\op{T}$ can be expressed in terms of transformations of $\Psi$ and $\vec{d}$ via\change{
\begin{align}
\op{g} \Psi(\vec{k}) &= \Psi\left(\mat{R}^{-1}(g)\vec{k}\right),\\
\op{T} \Psi(\vec{k}) &= \Psi(-\vec{k}),
\end{align}}
and\change{
\begin{align}
\op{g} \vec{d}(\vec{k}) &= \det\left(\mat{R}(g)\right)\mat{R}(g)\vec{d}\left(\mat{R}^{-1}(g)\vec{k}\right),\\
\op{T} \vec{d}(\vec{k})&= -\vec{d}(-\vec{k}).
\end{align}}
The gap function has to be odd under the application of a combination of the parity operator ($\op{P}$), spin interchange ($\op{S}$), and \change{relative time reversal}  ($\op{T}$) \cite{Triola2016},
\begin{equation}
\op{P}\op{S}\op{T} = -1.
\label{ist}
\end{equation}
Therefore, by considering an even behavior under time-reversal $\op{T}\mat{\Delta} = \mat{\Delta}$, a spin singlet gap (odd under spin interchange) restricts the gap function to be even under parity, whereas a spin triplet gap (even under spin interchange) has to come with an odd parity. We now know, by allowing for an odd-time (or odd-frequency) dependence of the gap function, $\op{T}\mat{\Delta} = -\mat{\Delta}$, that options for constructing an odd-parity spin singlet and an even-parity spin triplet gap arise. We proceed to show how the linearized gap equation allows for odd-frequency solutions.
\section{Shubnikov point groups}
\begin{figure}[!b]
\subfloat[first kind]{\includegraphics[height=2.6cm]{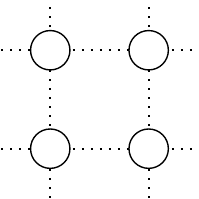}}\hspace{0.1cm}
\subfloat[second kind]{\includegraphics[height=2.6cm]{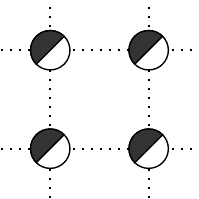}}\hspace{0.1cm}
\subfloat[third kind]{\includegraphics[height=2.6cm]{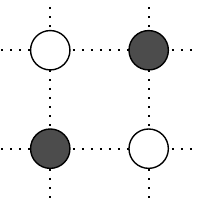}}
\caption{Illustration of the Shubnikov point-group construction.\label{fig:shub}}
\end{figure}
Superconductivity is mediated by a pairing of electrons in $\vec{k}$ space. In three dimensions it is possible to define seven crystal systems and 32 crystal classes. The latter are connected to the 32 point groups. Whereas point groups only describe the spatial symmetries of the system they can be extended to incorporate time-reversal symmetry. 

According to Eq. \eqref{tr_gap}, \change{relative time reversal} is a symmetry element of order 2, i.e., $\op{T}^2=\op{1}$. To include $\op{T}$ as a symmetry element, we follow the Shubnikov construction for colored groups. \change{However, note that the relative time-reversal considered here is a unitary symmetry element.} Denoting the point group of the system by $\mathcal{G}$, three kinds of Shubnikov groups can be defined (Fig. \ref{fig:shub}). The Shubnikov group of the first kind is given by the point group itself,
\begin{equation}
\mathcal{G}^{\text{I}} = \mathcal{G}.
\label{schub1}
\end{equation}
Shubnikov groups of the second kind are introduced by combining each element of $\mathcal{G}$ with $\op{T}$,
\begin{equation}
\mathcal{G}^{\text{II}} = \mathcal{G} + \op{T}\mathcal{G}.
\label{schub}
\end{equation}
Since each symmetry element occurs twice, once in connection to time-reversal and once without, these groups are also referred to as gray groups. 
Shubnikov groups of the third kind describe systems without a global time-reversal symmetry, e.g., magnetic structures. Here, time-reversal symmetry is only connected to a few elements. Starting from an invariant subgroup $\mathcal{N}\subset\mathcal{G}$ of index 2 ($\operatorname{ord}{\mathcal{G}/\mathcal{N}}=2$), Shubnikov groups of the third kind are constructed via
\begin{equation}
\mathcal{G}^{\text{III}} = \mathcal{N} + \op{T}(\mathcal{G}-\mathcal{N}).
\label{schub3}
\end{equation}
If the pairing potential in Eq. \eqref{gap_equation} is time-reversal symmetric, it is invariant under a group $\mathcal{G}^{\text{II}}$. Hence, to discuss odd- and even-frequency gaps for time-reversal symmetric pairing potentials, Shubnikov groups of the second kind are considered. In comparison to Shubnikov groups of the first kind and Shubnikov groups of the third kind, this approach allows for an explicit distinction between representations odd and even under \change{relative time reversal}, as will be explained within the next section.  

For Shubnikov groups of the second kind it follows from Eq. \eqref{schub} that $\operatorname{ord}{\mathcal{G}^{\text{II}}} = 2\operatorname{ord}\mathcal{G}$. Furthermore, $\op{T}$ commutes with every element of $\mathcal{G}^{\text{II}}$. Thus, $\left\{E,\op{T}\right\}$ is an Abelian invariant subgroup. It follows that $\mathcal{G}^{\text{II}}$ can be written as a semi-direct product of $\mathcal{G}$ and $\left\{E,\op{T}\right\}$, and, by induction \cite{hergert2018}, that twice as many irreducible representations occur for $\mathcal{G}^{\text{II}}$ in comparison to $\mathcal{G}$. If $\Gamma_i$ is an irreducible representation of $\mathcal{G}$, then $\Gamma_i^+$ and $\Gamma_i^-$ are irreducible representations of $\mathcal{G}^{\text{II}}$, where the characters are given by
\begin{align}
 \chi_i^+(\op{T} g) &= \chi_i(g),\label{trchi1} \\
 \chi_i^-(\op{T} g) &= -\chi_i(g),\label{trchi2} 
\end{align}
for all $g \in \mathcal{G}^{\text{II}}$.

\begin{table*}[t!]
\begin{tikzpicture}
\draw[] (0,0) node{
\begin{tabular}{c cccccccccc|cccccccccc}
\hline\hline\\[-2ex]
  & $E$ & $2 C_2'$ & $2\sigma_v$ & $2C_2''$ & $2 \sigma_d$ & $2S_4$ & $2C_4$ & $I$ & $C_2$ & $\sigma_h$ & $\op{T}$ & $2 \op{T}C_2'$ & $2\op{T}\sigma_v$ & $2\op{T}C_2''$ & $2 \op{T}\sigma_d$ & $2\op{T}S_4$ & $2\op{T}C_4$ & $\op{T}I$ & $\op{T}C_2$ & $\op{T}\sigma_h$\\
  \hline
 $\text{A}^{+}_{\text{1g}}$ & 1 & 1 & 1 & 1 & 1 & 1 & 1 & 1 & 1 & 1& 1 & 1 & 1 & 1 & 1 & 1 & 1 & 1 & 1 & 1 \\
 $\text{A}^{+}_{\text{2g}}$ & 1 & -1 & -1 & -1 & -1 & 1 & 1 & 1 & 1 & 1 & 1 & -1 & -1 & -1 & -1 & 1 & 1 & 1 & 1 & 1\\
 $\text{B}^{+}_{\text{1g}}$ & 1 & 1 & 1 & -1 & -1 & -1 & -1 & 1 & 1 & 1 & 1 & 1 & 1 & -1 & -1 & -1 & -1 & 1 & 1 & 1\\
 $\text{B}^{+}_{\text{2g}}$ & 1 & -1 & -1 & 1 & 1 & -1 & -1 & 1 & 1 & 1 & 1 & -1 & -1 & 1 & 1 & -1 & -1 & 1 & 1 & 1\\
 $\text{E}^{+}_{\text{g}}$ & 2 & 0 & 0 & 0 & 0 & 0 & 0 & 2 & -2 & -2 & 2 & 0 & 0 & 0 & 0 & 0 & 0 & 2 & -2 & -2 \\
 $\text{A}^{+}_{\text{1u}}$ & 1 & 1 & -1 & 1 & -1 & -1 & 1 & -1 & 1 & -1 & 1 & 1 & -1 & 1 & -1 & -1 & 1 & -1 & 1 & -1\\
 $\text{A}^{+}_{\text{2u}}$ & 1 & -1 & 1 & -1 & 1 & -1 & 1 & -1 & 1 & -1& 1 & -1 & 1 & -1 & 1 & -1 & 1 & -1 & 1 & -1 \\
 $\text{B}^{+}_{\text{1u}}$ & 1 & 1 & -1 & -1 & 1 & 1 & -1 & -1 & 1 & -1 & 1 & 1 & -1 & -1 & 1 & 1 & -1 & -1 & 1 & -1 \\
 $\text{B}^{+}_{\text{2u}}$ & 1 & -1 & 1 & 1 & -1 & 1 & -1 & -1 & 1 & -1  & 1 & -1 & 1 & 1 & -1 & 1 & -1 & -1 & 1 & -1 \\
 $\text{E}^{+}_{\text{u}}$ & 2 & 0 & 0 & 0 & 0 & 0 & 0 & -2 & -2 & 2 & 2 & 0 & 0 & 0 & 0 & 0 & 0 & -2 & -2 & 2\\
  \hline
  $\text{A}^{-}_{\text{1g}}$ & 1 & 1 & 1 & 1 & 1 & 1 & 1 & 1 & 1 & 1  & -1 & -1 & -1 & -1 & -1 & -1 & -1 & -1 & -1 & -1\\
  $\text{A}^{-}_{\text{2g}}$ & 1 & -1 & -1 & -1 & -1 & 1 & 1 & 1 & 1 & 1  
& -1 & 1 & 1 & 1 & 1 & -1 & -1 & -1 & -1 & -1\\
  $\text{B}^{-}_{\text{1g}}$ & 1 & 1 & 1 & -1 & -1 & -1 & -1 & 1 & 1 & 1
& -1 & -1 & -1 & 1 & 1 & 1 & 1 & -1 & -1 & -1 \\
  $\text{B}^{-}_{\text{2g}}$ & 1 & -1 & -1 & 1 & 1 & -1 & -1 & 1 & 1 & 1
 & -1 & 1 & 1 & -1 & -1 & 1 & 1 & -1 & -1 & -1 \\
  $\text{E}^{-}_{\text{g}}$ & 2 & 0 & 0 & 0 & 0 & 0 & 0 & 2 & -2 & -2 
 & -2 & 0 & 0 & 0 & 0 & 0 & 0 & -2 & 2 & 2\\
  $\text{A}^{-}_{\text{1u}}$ & 1 & 1 & -1 & 1 & -1 & -1 & 1 & -1 & 1 & -1
& -1 & -1 & 1 & -1 & 1 & 1 & -1 & 1 & -1 & 1  \\
  $\text{A}^{-}_{\text{2u}}$ & 1 & -1 & 1 & -1 & 1 & -1 & 1 & -1 & 1 & -1
& -1 & 1 & -1 & 1 & -1 & 1 & -1 & 1 & -1 & 1 \\
  $\text{B}^{-}_{\text{1u}}$ & 1 & 1 & -1 & -1 & 1 & 1 & -1 & -1 & 1 & -1
& -1 & -1 & 1 & 1 & -1 & -1 & 1 & 1 & -1 & 1 \\
  $\text{B}^{-}_{\text{2u}}$ & 1 & -1 & 1 & 1 & -1 & 1 & -1 & -1 & 1 & -1
 & -1 & 1 & -1 & -1 & 1 & -1 & 1 & 1 & -1 & 1 \\
  $\text{E}^{-}_{\text{u}}$ & 2 & 0 & 0 & 0 & 0 & 0 & 0 & -2 & -2 & 2 
& -2 & 0 & 0 & 0 & 0 & 0 & 0 & 2 & 2 & -2\\
 \hline\hline
\end{tabular}
};
\draw [decoration={brace,amplitude=0.5em},decorate,ultra thick,gray]
         (7.7,3.6) -- (7.7,-0.1);
\draw [decoration={brace,amplitude=0.5em},decorate,ultra thick,gray]
         (7.7,-0.4) -- (7.7,-4.1);
\draw[] (8,1.8) node[right]{$\boldsymbol{\op{T}}$\textbf{-even}};
\draw[] (8,-2.2) node[right]{$\boldsymbol{\op{T}}$\textbf{-odd}};
\end{tikzpicture}
\caption{Character table of the Shubnikov group ${D}^{\text{II}}_{4h}$.\label{ct_D4h}}
\end{table*}

Since the linearized gap equation \eqref{gap_equation2} is an eigenvalue equation, where the operator $\op{V}$ is invariant under all transformations of a symmetry group $\mathcal{G}^{\text{II}}$, each eigenfunction transforms as one of the irreducible representations. Hence, each eigenfunction or gap function is an eigenfunction of the character projection operator $\op{\mathcal{P}}^p$, given by
\begin{equation}
 \op{\mathcal{P}}^p = \sum_{g\in\mathcal{G}}\left(\chi^p(g)\right)^* \op{g}.
\end{equation}
Here, $\chi^p(g)$ denotes the character (the trace of the representation matrix) of the element $g$ within the irreducible representation $\Gamma^p$ of $\mathcal{G}$. Due to the orthogonality of irreducible representations $\op{\mathcal{P}}^p$ has the property
\begin{equation}
 \op{\mathcal{P}}^q \mat{\Delta}_{m}^{p,\nu} = \delta_{pq}\,\mat{\Delta}_{m}^{p,\nu}.
\end{equation}
Applied to an arbitrary gap function $\mat{\Delta}$, the character projection operator $\op{\mathcal{P}}^p$ projects out the part of $\mat{\Delta}$ transforming as the irreducible representation $\Gamma^p$, denoted by $\mat{\Delta}^p$,
\begin{equation}
 \op{\mathcal{P}}^p \mat{\Delta} = \sum_{\nu}\sum_{m=1}^{d_p} \mat{\Delta}_{m}^{p,\nu} = \mat{\Delta}^{p}.
\end{equation}
Taking into account all $N$ inequivalent irreducible representations $\Gamma^p$ of a symmetry group $\mathcal{G}$ and summing over all $\Delta^p$ obtained by mutual application of $\op{\mathcal{P}}^p$ to $\Delta$ the original gap function has to be revealed,
\begin{equation}
 \mat{\Delta} = \sum_{p=1}^{N}\mat{\Delta}^{p}.
\end{equation}
With these remarks we are now ready to analyze the self-consistent solution for gap functions. 
\section{Computational details}
The group theoretical analysis was performed by applying the \textsc{mathematica} group theory package GTPack \cite{hergert2018,HergertGeilhufe} (\url{http://gtpack.org}). As a faithful representation of the point-group elements, rotation matrices of the group $O(3)$ are used. Since $\mathcal{G}^{II}$ from Eq. \eqref{schub} is isomorphic to the direct product group $\mathcal{G}\otimes\left\{1,-1\right\}$, a faithful representation for the Shubnikov point group is found by the $4\times 4$-matrices
\begin{equation}
\mat{D}(g) = \left(\begin{array}{cc}
\mat{R}(g) & \mat{0} \\
\mat{0} & 1
\end{array}
\right),\quad \text{and}\quad 
\mat{D}(\op{T}g) = \left(\begin{array}{cc}
\mat{R}(g) & \mat{0} \\
\mat{0} & -1
\end{array}
\right).
\end{equation}
Character tables were calculated by applying the Burnside algorithm \cite{holt2005} which is a reasonable choice due to the small order of the crystallographic point groups. For the generalized Ginzburg-Landau theory, representation matrices of the irreducible representations and the corresponding Clebsch-Gordan coefficients were calculated by applying the algorithm of Flodmark and Blokker \cite{flodmark1967computer} and van Den Broek and Cornwell \cite{van1978clebsch}, respectively. The superconducting gap can be expanded in terms of tesseral harmonics $S^l_m$ (real spherical harmonics) as 
\begin{equation}
 \Psi(\vec{k}) = \sum_{l}\sum_{m=-l}^l c^l_m S^l_m(x,y,z),
 \label{psi:sin}
\end{equation}
and
\begin{equation}
 \vec{d}(\vec{k}) = \sum_{l}\sum_{m=-l}^l \vec{d}^l_m S^l_m(x,y,z).
 \label{d:trip}
\end{equation}
Throughout the paper the $S^l_m$ are discussed in Cartesian form.

\section{Square lattice ($D^{\text {II}}_{4h}$)}
To give a specific example of the emergence of even- and odd-frequency superconducting states we choose a specific group. In the following, a square lattice having the point group $D_{4h}$ is discussed. The group is generated by the elements $\left\{C_{4z},C_{2y},I\right\}$, where $C_{4z}$ denotes a four-fold rotation about the $z$ axis, $C_{2y}$ denotes a two-fold rotation about the $y$ axis and $I$ denotes the inversion. In total, $D_{4h}$ has 16 elements. Consequently, the corresponding Shubnikov group of the second kind ${D}^{\text{II}}_{4h}$ has 32 elements and is constructed according to Eq. \eqref{schub}. The character table of ${D}^{\text{II}}_{4h}$ is shown in Table \ref{ct_D4h}. For the irreducible representations the Mulliken notation is used \cite{Mulliken1955,mulliken1956}. Additionally, they are labeled with a superscript indicating an even (+) or odd (-) behavior with respect to time-reversal according to Eqs. \eqref{trchi1} and \eqref{trchi2}. 

For the spin singlet gaps, the allowed irreducible representations occurring for a certain angular momentum $l$ can be determined by decomposing the representations of the orbital part only (see Appendix A). These are given by
\begin{align}
 \text{$s$-wave}:\,D^0_{g,+} &\simeq \text{A}_{\text{1g}}^+, \label{d4h_res1}\\ 
 \text{$p$-wave}:\,D^1_{u,-} &\simeq \text{A}_{\text{2u}}^- \oplus \text{E}_{\text{u}}^-,\label{d4h_res2} \\ 
 \text{$d$-wave}:\,D^2_{g,+} &\simeq \text{A}_{\text{1g}}^+ \oplus \text{B}_{\text{1g}}^+ \oplus \text{B}_{\text{2g}}^+ \oplus \text{E}_{\text{g}}^+.\label{d4h_res3}
\end{align}
Analogously, for the spin triplet gaps the allowed irreducible representations are found by decomposing the direct product belonging to the orbital part with $D^1_{g,-}$, representing the transformation properties of the spin triplet state,
\begin{align}
 \text{$s$-wave}:\,D^0_{g,+}  \otimes D^1_{g,-} &\simeq \text{A}_{\text{2g}}^- \oplus \text{E}_{\text{g}}^-, \label{d4h_res4}\\ 
 \text{$p$-wave}:\,D^1_{u,-} \otimes D^1_{g,-} &\simeq \text{A}_{\text{2u}}^+ \oplus \text{B}_{\text{2u}}^+ \oplus \text{B}_{\text{1u}}^+\oplus 2 \text{A}_{\text{1u}}^+ \oplus 2 \text{E}_{\text{u}}^+,\label{d4h_res5} \\ 
\text{$d$-wave}:\,D^2_{g,+} \otimes D^1_{g,-} &\simeq \text{A}_{\text{1g}}^- \oplus 2\text{A}_{\text{2g}}^- \oplus 2\text{B}_{\text{1g}}^- \oplus 2\text{B}_{\text{2g}}^- \oplus 4\text{E}_{\text{g}}^-.\label{d4h_res6}
\end{align}
\begin{table}[!t]
\begin{tabular}{lll}
\hline
\hline
\multicolumn{3}{l}{even-frequency}\\
\hline
$s$-wave: & $\text{A}^{+}_{\text{1g}}$ & $\Psi \simeq \text{const},\,k_x^2+k_y^2+k_z^2$\\
$p$-wave: & $\text{A}^{+}_{\text{1u}}$ & $\vec{d}\simeq k_x\vec{e}_x+k_y\vec{e}_y+k_z\vec{e}_z$\\
& $\text{A}^{+}_{\text{1u}}$ & $\vec{d}\simeq 2k_z\vec{e}_z-k_x\vec{e}_x-k_y\vec{e}_y$\\
& $\text{A}^{+}_{\text{2u}}$ & $\vec{d}\simeq k_y\vec{e}_x-k_x\vec{e}_y$\\
& $\text{B}^{+}_{\text{1u}}$ & $\vec{d}\simeq k_x\vec{e}_x-k_y\vec{e}_y$\\
& $\text{B}^{+}_{\text{2u}}$ & $\vec{d}\simeq k_y\vec{e}_x+k_x\vec{e}_y$\\
& $\text{E}^{+}_{\text{u}}$ & $\vec{d}\simeq k_x\vec{e}_z$\\
&  & $\vec{d}\simeq k_y\vec{e}_z$\\
& $\text{E}^{+}_{\text{u}}$ & $\vec{d}\simeq k_z\vec{e}_x$\\
&  & $\vec{d}\simeq k_z\vec{e}_y$\\
$d$-wave: & $\text{A}^{+}_{\text{1g}}$ & $\Psi \simeq 2k_z^2-k_x^2-k_y^2$\\
& $\text{B}^{+}_{\text{1g}}$ & $\Psi \simeq (k_x^2-k_y^2)$\\
& $\text{B}^{+}_{\text{2g}}$ & $\Psi \simeq k_x k_y$\\
& $\text{E}^{+}_{\text{g}}$ & $\Psi \simeq k_x k_z$\\
& & $\Psi \simeq k_y k_z$\\
\hline
\multicolumn{3}{l}{odd-frequency}\\
\hline
$s$-wave: &  $\text{A}^{-}_{\text{2g}}$ & $\vec{d}\simeq (k_x^2+k_y^2+k_z^2)\vec{e}_z$\\
 &  $\text{E}^{-}_{\text{g}}$ & $\vec{d}\simeq (k_x^2+k_y^2+k_z^2)\vec{e}_x$\\
 &    & $\vec{d}\simeq (k_x^2+k_y^2+k_z^2)\vec{e}_y$\\
$p$-wave: &  $\text{A}^{-}_{\text{2u}}$ & $\Psi \simeq k_z$ \\
  &  $\text{E}^{-}_{\text{u}}$ & $\Psi \simeq k_x$ \\
  &   & $\Psi \simeq k_y$ \\
$d$-wave: &  $\text{A}^{-}_{\text{1g}}$ & $\vec{d}\simeq k_y k_z \vec{e}_x - k_x k_z \vec{e}_y$ \\
 &  $\text{A}^{-}_{\text{2g}}$ & $\vec{d}\simeq k_x k_z \vec{e}_x + k_y k_z \vec{e}_y$ \\
 &  $\text{A}^{-}_{\text{2g}}$ & $\vec{d}\simeq (2 k_z^2 - k_x^2 - k_y^2) \vec{e}_z$ \\

 &  $\text{B}^{-}_{\text{1g}}$ & $\vec{d}\simeq k_y k_z \vec{e}_x + k_x k_z \vec{e}_y$ \\
 &  $\text{B}^{-}_{\text{1g}}$ & $\vec{d}\simeq k_x k_y \vec{e}_z$ \\
 &  $\text{B}^{-}_{\text{2g}}$ & $\vec{d}\simeq k_x k_z \vec{e}_x - k_y k_z \vec{e}_y$ \\
 &  $\text{B}^{-}_{\text{2g}}$ & $\vec{d}\simeq (k_x^2- k_y^2) \vec{e}_z$ \\

 &  $\text{E}^{-}_{\text{g}}$ & $\vec{d}\simeq k_x k_y \vec{e}_x$ \\
 &   & $\vec{d}\simeq k_x k_y \vec{e}_y$ \\
 
 &  $\text{E}^{-}_{\text{g}}$ & $\vec{d}\simeq k_z k_y \vec{e}_z$ \\
 &   & $\vec{d}\simeq k_z k_x \vec{e}_z$ \\

 &  $\text{E}^{-}_{\text{g}}$ & $\vec{d}\simeq (2 k_z^2 - k_x^2 - k_y^2) \vec{e}_x$ \\
 & & $\vec{d}\simeq (k_x^2 - k_y^2) \vec{e}_x$ \\
 &  $\text{E}^{-}_{\text{g}}$ & $\vec{d}\simeq (2 k_z^2 - k_x^2 - k_y^2) \vec{e}_y$ \\
 & & $\vec{d}\simeq (k_x^2 - k_y^2) \vec{e}_y$ \\
\hline
\hline
\end{tabular}
\caption{Even- and odd-frequency gap symmetries for the square lattice (${D}^{\text{II}}_{4h}$), considering $s$-, $p$- and $d$-wave superconductivity.\label{gap_sym}}
\end{table}
The obtained terms in Eqs. \eqref{d4h_res1}-\eqref{d4h_res6} are in agreement with Eq. \eqref{ist}. They reflect the options:
\begin{itemize}
\item spin singlet, even parity, even time: \eqref{d4h_res1} and \eqref{d4h_res3} 
\item spin singlet, odd parity, odd time: \eqref{d4h_res2}
\item spin triplet, odd parity, even time: \eqref{d4h_res5}
\item spin-triplet, even parity, odd time: \eqref{d4h_res4} and \eqref{d4h_res6}
\end{itemize}
Specific terms for gap symmetries are obtained by applying the character projection operator to Eqs. \eqref{psi:sin} and \eqref{d:trip}. The results are illustrated in Table \ref{ct_D4h} and discussed subsequently for two examples.

\subsection{$s$-wave spin triplet}
\begin{figure}[b!]
\subfloat[spin triplet, $s$-wave\label{fig:trip:s:d4h}]{\includegraphics[height=3cm]{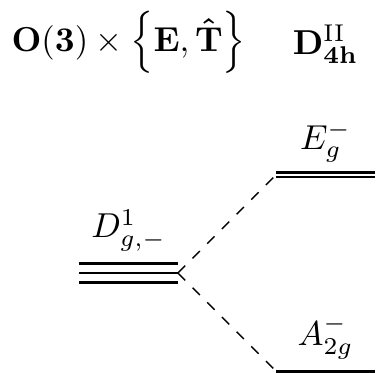}}\hspace{1cm}
\subfloat[spin singlet, $p$-wave\label{fig:sin:p:d4h}]{\includegraphics[height=3cm]{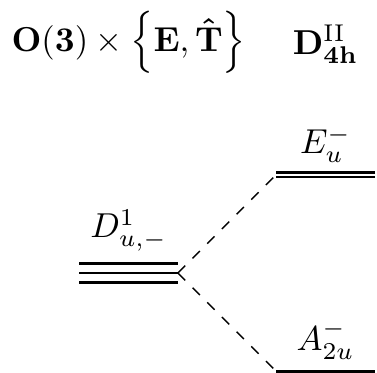}}
\caption{Splitting of pairing states for a pairing potential with $D_{4h}^{\text{II}}$ symmetry.}
\end{figure}
As a first example, we consider a $s$-wave superconductor. Whereas the conventional BCS theory \cite{bardeen1957theory} describes a $s$-wave spin singlet pairing which is even under time-reversal, it is possible to construct a $s$-wave spin triplet that is odd under time-reversal [Eq. \eqref{d4h_res4}]. Under full rotational symmetry, a spin triplet transforms as the three-dimensional representation $D_{g,-}^1$ as discussed in Appendix A. However, for the square lattice, the triplet state splits into $\text{A}_{\text{2g}}^-$ and $\text{E}_{\text{g}}^-$ as illustrated in Figure \ref{fig:trip:s:d4h}. Since the $z$-axis is chosen as the principal axis, two linearly independent solutions belonging to $\text{E}_{\text{g}}^-$ transform as $\vec{k}^2\vec{e}_x$ and $\vec{k}^2\vec{e}_y$. Solutions belonging to $\text{A}_{\text{2g}}^-$ transform as $\vec{k}^2\vec{e}_z$. The resulting gap functions are given by
\begin{align}
  \mat{\Delta}^{\text{E}_{\text{g}}^-}_1(\vec{k}) &= -\vec{k}^2\mat{\sigma}_z,\label{d4h:s:trip:1}\\
  \mat{\Delta}^{\text{E}_{\text{g}}^-}_2(\vec{k}) &= i \vec{k}^2 \mat{\sigma}_0,\label{d4h:s:trip:2}
\end{align}
and
\begin{equation}
  \mat{\Delta}^{\text{A}_{\text{2g}}^-}_1(\vec{k}) = \vec{k}^2\mat{\sigma}_x.\label{d4h:s:trip:3}
\end{equation}
As expected, all three matrices are symmetric and thus even under spin interchange. They are even under parity since they contain $\vec{k}^2$. But, they are odd with respect to the \change{relative time reversal}  introduced in Eq. \eqref{tr_gap}.
\subsection{$p$-wave spin singlet}
Another unconventional odd-frequency pairing is given by the $p$-wave spin singlet. Here, the three-dimensional odd-parity representation $D_{u,-}^1$ splits into the irreducible representations $\text{A}_{\text{2u}}^-$ and $\text{E}_{\text{u}}^-$. The gap transforms as $k_x$ and $k_y$ for $\text{E}_{\text{u}}^-$ and as $k_z$ for $\text{A}_{\text{2u}}^-$. The resulting superconducting gaps behave as
\begin{align}
  \mat{\Delta}^{\text{E}_{\text{u}}^-}_1(\vec{k}) &= i k_x\mat{\sigma}_y,\\
  \mat{\Delta}^{\text{E}_{\text{u}}^-}_2(\vec{k}) &= i k_y \mat{\sigma}_y,
\end{align}
and
\begin{equation}
  \mat{\Delta}^{\text{A}_{\text{2u}}^-}_1(\vec{k}) = i k_z \mat{\sigma}_y.
\end{equation}
Clearly, the three matrices are anti-symmetric and odd under spin, odd under parity, and also odd under \change{relative time reversal}  according to Eq. \eqref{tr_gap}.
\begin{table}[h!]
\begin{tabular}{lll}
\hline
\hline
\multicolumn{3}{l}{even-frequency}\\
\hline
$s$-wave: & $\text{A}^{+}_{\text{1g}}$ & $\Psi \simeq \text{const},\,k_x^2+k_y^2+k_z^2$\\
$p$-wave: & $\text{A}^{+}_{\text{1u}}$ & $\vec{d} \simeq k_x \vec{e}_x + k_y \vec{e}_y + k_z \vec{e}_z$\\
& $\text{E}^{+}_{\text{u}}$ & $\vec{d} \simeq k_x \vec{e}_x - k_y \vec{e}_y$ \\
 & & $\vec{d} \simeq 2 k_z \vec{e}_z - k_x \vec{e}_x - k_y \vec{e}_y$ \\
& $\text{T}^{+}_{\text{1u}}$ & $\vec{d} \simeq k_y \vec{e}_x - k_x \vec{e}_y$ \\
 & & $\vec{d} \simeq k_z \vec{e}_y - k_y \vec{e}_z$ \\
 & & $\vec{d} \simeq k_x \vec{e}_z - k_z \vec{e}_x$ \\
& $\text{T}^{+}_{\text{2u}}$ & $\vec{d} \simeq k_y \vec{e}_x + k_x \vec{e}_y$ \\
 & & $\vec{d} \simeq k_z \vec{e}_y + k_y \vec{e}_z$ \\
 & & $\vec{d} \simeq k_x \vec{e}_z + k_z \vec{e}_x$ \\
$d$-wave: & $\text{E}^{+}_{\text{g}}$ & $\Psi \simeq (k_x^2-k_y^2)$\\
  &   & $\Psi \simeq (2 k_z^2-k_x^2-k_y^2)$\\
& $\text{T}^{+}_{\text{2g}}$ & $\Psi \simeq k_x k_y$\\
  &   & $\Psi \simeq k_y k_z$\\
  &   & $\Psi \simeq k_x k_z$\\
\hline
\multicolumn{3}{l}{odd-frequency}\\
\hline
$s$-wave: & $\text{T}^{-}_{\text{1g}}$ & $\vec{d} \simeq (k_x^2+k_y^2+k_z^2)\vec{e}_x$\\
& & $\vec{d} \simeq (k_x^2+k_y^2+k_z^2)\vec{e}_y$\\
& & $\vec{d} \simeq (k_x^2+k_y^2+k_z^2)\vec{e}_z$\\
$p$-wave: & $\text{T}^{-}_{\text{1u}}$ & $\Psi \simeq k_x$\\
& & $\Psi \simeq k_y$\\
& & $\Psi \simeq k_z$\\
$d$-wave: & $\text{A}^{-}_{\text{2g}}$ & $\vec{d} \simeq k_y k_z \vec{e}_x + k_x k_z \vec{e}_y + k_x k_y \vec{e}_z$\\
& $\text{E}^{-}_{\text{g}}$ & $\vec{d} \simeq k_y k_z \vec{e}_x - k_x k_z \vec{e}_y$ \\
 & & $\vec{d} \simeq 2 k_x k_y \vec{e}_z - k_y k_z \vec{e}_x - k_x k_z \vec{e}_y$ \\
 & $\text{T}^{-}_{\text{1g}}$ & $\vec{d} \simeq k_x k_y \vec{e}_x + k_y k_z \vec{e}_z$ \\
 & & $\vec{d} \simeq k_x k_z \vec{e}_x + k_y k_z \vec{e}_y$ \\
 & & $\vec{d} \simeq k_x k_y \vec{e}_y + k_x k_z \vec{e}_z$ \\
 
 & $\text{T}^{-}_{\text{1g}}$ & $\vec{d} \simeq (2 k_x^2 - k_y^2 - k_z^2) \vec{e}_x$ \\
 & & $\vec{d} \simeq (2 k_y^2 - k_z^2 - k_x^2) \vec{e}_y$ \\
 & & $\vec{d} \simeq (2 k_z^2 - k_x^2 - k_y^2) \vec{e}_z$ \\ 
 
  & $\text{T}^{-}_{\text{2g}}$ & $\vec{d} \simeq k_x k_y \vec{e}_x - k_y k_z \vec{e}_z$ \\
 & & $\vec{d} \simeq k_x k_z \vec{e}_x - k_y k_z \vec{e}_y$ \\
 & & $\vec{d} \simeq k_x k_y \vec{e}_y - k_x k_z \vec{e}_z$ \\

 & $\text{T}^{-}_{\text{2g}}$ & $\vec{d} \simeq (k_y^2 - k_z^2) \vec{e}_x$ \\
 & & $\vec{d} \simeq (k_z^2 - k_x^2) \vec{e}_y$ \\
 & & $\vec{d} \simeq (k_x^2 - k_y^2) \vec{e}_z$ \\ 
\hline
\hline
\end{tabular}
\caption{Even- and odd-frequency gap symmetries for cubic lattices with octahedral symmetry ($O_{h}$), considering $s$-, $p$- and $d$-wave superconductivity.\label{gap_sym_oh}}
\end{table}

\section{Octahedral symmetry ($O^{\text {II}}_{h}$)}
Similarly to $D^{\text {II}}_{4h}$,the gap symmetry is analyzed for $O^{\text {II}}_{h}$. As mentioned before, due to the semi-direct product structure of Shubnikov groups of the second kind each irreducible representation of a point group occurs twice (see Table \ref{ct_D4h} for $D^{\text {II}}_{4h}$). Therefore, we stick to the standard nomenclature of the irreducible representations of $O_h$ as can be found, e.g., in Refs. \cite{hergert2018,altmann1994point,cornwell1984group}. As in the example of $D^{\text {II}}_{4h}$, superscripts $+$ and $-$ distinguish between the even and odd symmetric representations with respect to \change{relative time reversal} . For $O^{\text {II}}_{h}$, the occurring irreducible representations for spin singlet and triplet gaps are given by
\begin{align}
\text{$s$-wave}:\,D^0_{g,+} &\simeq \text{A}_{\text{1g}}^+, \\ 
\text{$p$-wave}:\,D^1_{u,-}&\simeq \text{T}_{\text{1u}}^-, \\ 
\text{$d$-wave}:\,D^2_{g,+} &\simeq \text{E}_{\text{g}}^+ \oplus \text{T}_{\text{2g}}^+,
\end{align}
and
\begin{align}
 \text{$s$-wave}:\,D^0_{g,+}  \otimes D^1_{g,-} &\simeq \text{T}_{\text{1g}}^-, \\ 
 \text{$p$-wave}:\,D^1_{u,-}  \otimes D^1_{g,-} &\simeq \text{A}_{\text{1u}}^+ \oplus \text{E}_{\text{u}}^+ \oplus \text{T}_{\text{1u}}^+\oplus  \text{T}_{\text{2u}}^+, \\ 
 \text{$d$-wave}:\,D^2_{g,+}  \otimes D^1_{g,-} &\simeq \text{A}_{\text{2g}}^- \oplus \text{E}_{\text{g}}^- \oplus 2\text{T}_{\text{1g}}^- \oplus 2\text{T}_{\text{2g}}^-,\label{oh_res6}
\end{align}
respectively. The specific forms of the even- and odd-frequency gap symmetries are shown in Table \ref{gap_sym_oh}. 
\subsection{$s$-wave spin triplet}
\begin{figure}[t!]
\includegraphics[height=3cm]{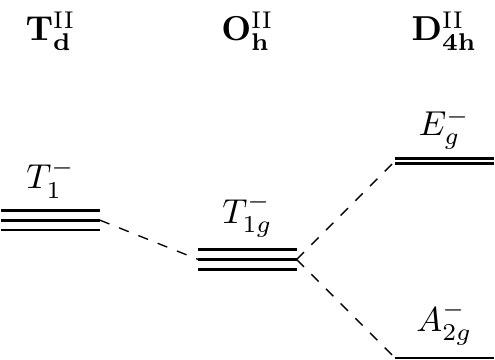}
\caption{$s$-wave spin triplet pairing for a pairing potential with $T^{\text {II}}_{d}$, $O^{\text {II}}_{h}$ and $D^{\text {II}}_{4h}$ symmetry.\label{s:sing:oh}}
\end{figure}

For octahedral symmetry the three-dimensional representation $D^1_{g,-}$ does not split and thus a three-fold degenerate eigenvalue occurs transforming as the irreducible representation $\text{T}_{\text{1g}}^-$. Similarly to Eqs. \eqref{d4h:s:trip:1}-\eqref{d4h:s:trip:3}, the resulting gaps transform as
\begin{align}
  \mat{\Delta}^{\text{T}_{\text{1g}}^-}_1(\vec{k}) &= -\vec{k}^2\mat{\sigma}_z,\\
  \mat{\Delta}^{\text{T}_{\text{1g}}^-}_2(\vec{k}) &= i \vec{k}^2 \mat{\sigma}_0,\\
  \mat{\Delta}^{\text{T}_{\text{1g}}^-}_3(\vec{k}) &= \vec{k}^2\mat{\sigma}_x.
\end{align}
The three matrices are even under spin interchange, even under parity and odd under \change{relative time reversal} . The relationship between the $s$-wave spin triplets for $O^{\text {II}}_{h}$ and $D^{\text {II}}_{4h}$ symmetry is shown in Fig. \ref{s:sing:oh}.

\subsection{$d$-wave spin triplet}
Another option of having a spin triplet, and even parity, but odd time pairing is given by the $d$-wave spin triplet. According to Eq. \eqref{oh_res6} and Table \ref{gap_sym_oh}, the $d$-wave spin triplet pairing is mediated by 15 eigenvectors belonging to six subspaces. As an example we consider the two subspaces belonging to ${\text{T}_{\text{2g}}^-}$. The corresponding gaps transform as  
\begin{align}
 \mat{\Delta}^{\text{T}_{\text{2g}}^-}_1(\vec{k}) &= -k_x k_y \mat{\sigma}_z - k_y k_z \mat{\sigma}_x \\
  \mat{\Delta}^{\text{T}_{\text{2g}}^-}_2(\vec{k}) &= -k_x k_z \mat{\sigma}_z - i k_y k_z \mat{\sigma}_0 \\
\mat{\Delta}^{\text{T}_{\text{2g}}^-}_1(\vec{k}) &= i k_x k_y \mat{\sigma}_0 - k_x k_z \mat{\sigma}_x \\
\end{align}
and
\begin{align}
  \mat{\Delta}^{\text{T}_{\text{2g}}^-}_1(\vec{k}) &= -(k_y^2-k_z^2)\mat{\sigma}_z,\\
  \mat{\Delta}^{\text{T}_{\text{2g}}^-}_2(\vec{k}) &= i (k_z^2-k_x^2) \mat{\sigma}_0,\\
  \mat{\Delta}^{\text{T}_{\text{2g}}^-}_3(\vec{k}) &= (k_x^2-k_y^2)\mat{\sigma}_x.
\end{align}
Both sets of matrices belong to two different eigenvalues within the linearized gap equation \eqref{gap_equation2}. Due to the absence of $\mat{\sigma}_y$, all matrices are symmetric and thus represent a gap even under spin interchange. Second order terms in $\vec{k}$ guarantee even behavior under parity. Nevertheless, following Eq. \eqref{tr_gap} they are odd under \change{relative time reversal} . 

\section{Non-centrosymmetric tetragonal lattice ($T_d^{\text{II}}$)}
Even for non-centrosymmetric groups, i.e., groups that do not contain the inversion, it is possible to keep the notion of parity within the gap function. The point group $T_d$ describes the point group of a tetragonal lattice having no inversion symmetry. However, as a subgroup of $O_h$, the irreducible representations of both groups can be related to each other. In the context of the Shubnikov group of the second kind $T_d^{\text{II}}\subset O_h^{\text{II}}$, the following correspondences can be found, 
\begin{align}
\text{A}^\pm_{\text{1g}}, \text{A}^\pm_{\text{2u}} &\rightarrow \text{A}^\pm_{\text{1}},\label{OfToTd1} \\
\text{A}^\pm_{\text{2g}}, \text{A}^\pm_{\text{1u}} &\rightarrow \text{A}^\pm_{\text{2}},\label{OfToTd2} \\
\text{E}^\pm_{\text{u}}, \text{E}^\pm_{\text{g}} &\rightarrow \text{E}^\pm,\label{OfToTd3} \\
\text{T}^\pm_{\text{1g}}, \text{T}^\pm_{\text{2u}} &\rightarrow \text{T}^\pm_{\text{1}},\label{OfToTd4} \\
\text{T}^\pm_{\text{2g}}, \text{T}^\pm_{\text{1u}} &\rightarrow \text{T}^\pm_{\text{2}}.\label{OfToTd5}
\end{align}
Hence, the specific gap symmetries for $T_d^{\text{II}}$ can be taken from Table \ref{gap_sym_oh}. The occurring irreducible representations are given by 
\begin{align}
\text{$s$-wave}:\,D^0_{g,+} &\simeq \text{A}_{\text{1}}^+, \\ 
\text{$p$-wave}:\,D^1_{u,-} &\simeq \text{T}_{\text{2}}^-, \\ 
\text{$d$-wave}:\,D^2_{g,+} &\simeq \text{E}^+ \oplus \text{T}_{\text{2}}^+,
\end{align}
and
\begin{align}
 \text{$s$-wave}:\,D^0_{g,+}  \otimes D^1_{g,-} &\simeq \text{T}_{\text{1}}^-, \\ 
 \text{$p$-wave}:\,D^1_{u,-}  \otimes D^1_{g,-} &\simeq \text{A}_{\text{2}}^+ \oplus \text{E}^+ \oplus \text{T}_{\text{2}}^+\oplus  \text{T}_{\text{1}}^+, \\ 
  \text{$d$-wave}:\,D^2_{g,+}  \otimes D^1_{g,-} &\simeq \text{A}_{\text{2}}^- \oplus \text{E}^- \oplus 2\text{T}_{\text{2}}^- \oplus 2\text{T}_{\text{1}}^-.
\end{align}
The discussion of examples follows the line of $O_h^{\text{II}}$ in the previous section. The relationship between $T_d^{\text{II}}$, $O_h^{\text{II}}$ and $D_{4h}^{\text{II}}$ for $s$-wave spin triplet pairing is shown in Fig. \ref{s:sing:oh}.

\section{Ginzburg-Landau theory}
The transition to a superconducting state occurs when the largest eigenvalue of the Bethe-Salpeter equation \eqref{gap_equation} is equal to $\nu=1$. This relation also uniquely defines the superconducting transition temperature $T_c$. The gap related to the first superconducting state arising immediately below $T_c$ can be constructed as a linear combination of the eigenfunctions of the Bethe-Salpeter equation $\mat{\Delta}^i_m$, $m=1,\dots,d_i$,
\begin{equation}
  \mat{\Delta}^i = \sum_{m=1}^{d_i} \eta_m \mat{\Delta}^i_m.
  \label{eq:gl:gap}
\end{equation}
The generalized Ginzburg-Landau free energy \cite{sigrist1991phenomenological} can be entirely expressed in terms of the $\eta_m$, 
\begin{equation}
 F(T,\vec{\eta}) = F_{\text{Landau}}(T,\vec{\eta}) + F_{\text{Gradient}}(\vec{\eta}) + \int \mathrm{d}x^3 \frac{\vec{B}^2}{8\pi}.
\end{equation}
We start the discussion with the Landau term, given by
\begin{multline}
  F_{\text{Landau}}(T,\vec{\eta}) = F_0(T) \\ + V \left[A_i(T)\sum_{m=1}^{d_i} \left|\eta_m\right|^2+f\left(\vec{\eta}^4\right)\right].
  \label{gl_free}
\end{multline}
Here, $f\left(\vec{\eta}^4\right)$ denotes all fourth-order terms in $\eta_m$ and its complex conjugate $\eta_m^*$. Since $F(T,\vec{\eta})$ has to be real, only products containing the same number of $\eta_m$ and $\eta_m^*$ are allowed. Furthermore, $F(T,\vec{\eta})$ has the same symmetry as the system itself and thus transforms as the identity representation.

Since the gap and the anomalous Green's function vanish at equal times (effectively $t=0$) the order parameter for odd-frequency superconductivity is widely discussed, e.g., by considering a composite order of a Cooper pair and a charge or spin fluctuation \cite{balatsky1994properties,abrahams1995properties,Dahal2009,linder2017odd}. Since the minimum of Eq. \eqref{gl_free} is achieved for a particular value of the time-independent vector $\vec{\eta}$, it can be regarded as a generalized order parameter covering both odd- and even-frequency superconductivity.

The forth-order terms in Eq. \eqref{gl_free} only depend on the dimension of the irreducible representation but not on the specific characteristics. The number of different invariant terms can be determined by decomposing the direct product $\left[\Gamma\right]^4=\Gamma^*\otimes\Gamma\otimes\Gamma^*\otimes\Gamma$. Here, we introduce the short-hand notation $\left[\Gamma\right]^4$ for convenience, e.g., $\left[\text{A}_{\text{1g}}^+\right]^4=(\text{A}_{\text{1g}}^+)^*\otimes\text{A}_{\text{1g}}^+\otimes(\text{A}_{\text{1g}}^+)^*\otimes\text{A}_{\text{1g}}^+$. For the groups $D_{4h}^{\text{II}}$ and $O_{h}^{\text{II}}$ it follows that
\begin{align}
\left[\text{A}_{\text{ix}}^\pm\right]^4 &\simeq \left[\text{B}_{\text{ix}}^\pm\right]^4 \simeq \text{A}_{\text{1g}}^+, \\
\left[\text{E}_{\text{x}}^\pm\right]^4 &\simeq 4\text{A}_{\text{1g}}^+ \oplus 4\text{A}_{\text{2g}}^+ \oplus 4\text{B}_{\text{1g}}^+ \oplus 4\text{B}_{\text{2g}}^+
\end{align}
  and
\begin{align}
\left[\text{A}_{\text{ix}}^\pm\right]^4 &\simeq \text{A}_{\text{1g}}^+, \\
\left[\text{E}_{\text{x}}^\pm\right]^4 &\simeq 3\text{A}_{\text{1g}}^+ \oplus 3\text{A}_{\text{2g}}^+ \oplus 5\text{E}_{\text{g}}^+,\\
\left[\text{T}_{\text{ix}}^\pm\right]^4 &\simeq 4\text{A}_{\text{1g}}^+ \oplus 3\text{A}_{\text{2g}}^+ \oplus 7\text{E}_{\text{g}}^+ \oplus 10\text{T}_{\text{1g}}^+ \oplus 10\text{T}_{\text{2g}}^+,
\end{align}
respectively. The abbreviations $i=1,2$ and $x=g,u$ were used. The analogous terms for $T_{d}^{\text{II}}$ can be derived using Eqs. \eqref{OfToTd1}-\eqref{OfToTd5}. 

The gradient term $F_{\text{Gradient}}$ of the free-energy incorporates a gauge-invariant coupling of the order
parameter to a magnetic field via the gradient vector $\vec{D}=\nabla - 2i\frac{e}{c} \vec{A}$. As $\vec{D}$ transforms as the vector representation $\Gamma_V$ of the underlying point group, the contributions of $F_{\text{Gradient}}$ are obtained from decomposing the direct product
\begin{equation}
 \Gamma_V^* \otimes \Gamma^{*} \otimes \Gamma_V\otimes \Gamma = \left[\Gamma\right]_V.
 \label{GL_GammaV}
\end{equation}
Similarly to $\left[\Gamma\right]^4$, $\left[\Gamma\right]_V$ is an abbreviation for the direct product in Eq. \eqref{GL_GammaV}, e.g., $\left[\text{A}_{\text{1g}}^+\right]_V=\left(\Gamma_V\right)^*\otimes\left(\text{A}_{\text{1g}}^+\right)^*\otimes\Gamma_V\otimes\text{A}_{\text{1g}}^+$. The vector representation of $D_{4h}^{\text{II}}$ is given by $\Gamma_V= \text{A}_{\text{1u}}^+ \oplus \text{E}_{\text{u}}^+$ and the vector representation of $O_{h}^{\text{II}}$ is $\Gamma_V= \text{T}_{\text{1u}}^+$. The corresponding Clebsch-Gordan sums for the decomposition of the gradient terms are
\begin{align}
\left[\text{A}_{\text{ix}}^\pm\right]_V &\simeq \left[\text{B}_{\text{ix}}^\pm\right]_V \simeq 2 \text{A}_{\text{1g}}^+ \oplus \text{A}_{\text{2g}}^+ \oplus \text{B}_{\text{1g}}^+ \oplus\text{B}_{\text{2g}}^+ \oplus 2 \text{E}_{\text{g}}^+,\label{fourth_rep_gl} \\
\left[\text{E}_{\text{x}}^\pm\right]_V &\simeq 5 \text{A}_{\text{1g}}^+ \oplus 5\text{A}_{\text{2g}}^+ \oplus 5\text{B}_{\text{1g}}^+ \oplus 5\text{B}_{\text{2g}}^+ \oplus  8\text{E}_{\text{g}}^+
\end{align}
for $D_{4h}^{\text{II}}$ and
\begin{align}
\left[\text{A}_{\text{ix}}^\pm\right]_V &\simeq \text{A}_{\text{1g}}^+ \oplus \text{E}_{\text{g}}^+ \oplus \text{T}_{\text{1g}}^+ \oplus \text{T}_{\text{2g}}^+, \\
\left[\text{E}_{\text{x}}^\pm\right]_V &\simeq 2\text{A}_{\text{1g}}^+ \oplus 2\text{A}_{\text{2g}}^+ \oplus 4\text{E}_{\text{g}}^+ \oplus 4\text{T}_{\text{1g}}^+ \oplus 4\text{T}_{\text{2g}}^+, \\
\left[\text{T}_{\text{ix}}^\pm\right]_V&\simeq 4\text{A}_{\text{1g}}^+ \oplus 3\text{A}_{\text{2g}}^+ \oplus 7\text{E}_{\text{g}}^+ \oplus 10\text{T}_{\text{1g}}^+ \oplus 10\text{T}_{\text{2g}}^+\label{vec_rep_gl}
\end{align}
for $O_{h}^{\text{II}}$. 

With respect to the second order, fourth order and gradient terms within the Ginzburg-Landau free energy functional \eqref{gl_free} it turns out that the explicit forms of the invariant polynomials only depend on the dimension of the irreducible representation involved. This statement follows from Eqs. \eqref{fourth_rep_gl}-\eqref{vec_rep_gl}. However, we want to exemplify the derivation of the invariant terms for the example of a tetragonal symmetry ($D_{4h}^{\text{II}}$) and an odd-frequency $p$-wave gap transforming as the irreducible representation $\text{E}_{\text{2u}}^-$. Hence, the superconducting gap is expressed as 
\begin{equation}
  \mat{\Delta}(\vec{k}) = \eta_1 \mat{\Delta}_1(\vec{k})+\eta_2 \mat{\Delta}_2(\vec{k}),
  \label{eq:gl:gapEu}
\end{equation}
where $\mat{\Delta}_1$ and $\mat{\Delta}_2$ transform as basis functions of $\text{E}_{\text{2u}}^-$. As $\left[\text{E}_{\text{2u}}^-\right]^4\simeq \left(\text{A}_{\text{1g}}^+\oplus\text{A}_{\text{2g}}^+\oplus\text{B}_{\text{1g}}^+\oplus\text{B}_{\text{2g}}^+\right)\otimes \left(\text{A}_{\text{1g}}^+\oplus\text{A}_{\text{2g}}^+\oplus\text{B}_{\text{1g}}^+\oplus\text{B}_{\text{2g}}^+\right)$, we start with forming a direct product basis in $\eta_i$ and $\eta^*_i$, $i=1,2$, for the direct product representations $\text{E}_{\text{2u}}^{-*}\otimes\text{E}_{\text{2u}}^-\simeq\text{A}_{\text{1g}}^+\oplus\text{A}_{\text{2g}}^+\oplus\text{B}_{\text{1g}}^+\oplus\text{B}_{\text{2g}}^+$. This can be done straightforwardly from the Clebsch-Gordan coefficients which were calculated using GTPack. Products of these basis functions span the invariant basis of the fourth-order terms. It turns out that only three independent fourth order polynomial terms remain in total, which are given by
\begin{multline}
 f\left(\vec{\eta}^4\right) = \beta_1 \left[\left|\eta_1\right|^4+\left|\eta_2\right|^4 \right]
 + \beta_2 \left[\eta_1^{2*}\eta_2^{2}+\text{h.c.} \right] \\+ \beta_3 \left[\left|\eta_1\right|^2\left|\eta_2\right|^2 \right].
\end{multline} 
These expressions are similar to the ones reported by Sigrist and Ueda \cite{sigrist1991phenomenological}, by setting $\beta_1=\beta_1'$, $\beta_2=4\beta_1'-4\beta_2'+\beta_3'$, and $\beta_3=\beta_3'$, where $\beta_i'$ denote the coefficients chosen in Ref. \cite{sigrist1991phenomenological}. For the gradient terms we follow a similar procedure as for the fourth-order terms. The basis of the vector representation is chosen to be $D_x$ and $D_y$ for $\text{E}_{\text{u}}^+$ and $D_z$ for $\text{A}_{\text{2u}}^+$. The resulting terms are 
\begin{multline}
 F_{\text{Gradient}} = \gamma_1\left[ \left|D_x\eta_1\right|^2 +  \left|D_y\eta_2\right|^2 \right] \\ + \gamma_2\left[ \left|D_y\eta_1\right|^2 +  \left|D_x\eta_2\right|^2  \right] + \gamma_3 \left[D_x\eta_1D_y^*\eta_2^* + \text{c.c.} \right] \\
 + \gamma_4 \left[D_x\eta_2D_y^*\eta_1^* + \text{c.c.} \right].
\end{multline}
Explicit expressions of fourth-order invariant and gradient terms for $D_{4h}$ and $O_{h}$ were reported before and discussed in great detail \cite{volovik1984unusual,volovik1985superconducting,blount1985symmetry,ueda1985p,sigrist1991phenomenological,sigrist1987symmetry}.

Additional terms resulting from external fields can be included into the Ginzburg-Landau theory leading to further contributions to the free-energy functional. With respect to odd-frequency superconductivity, fields which transform as an irreducible representation odd in time-reversal or Matsubara frequency, respectively, might be of special interest.

\section*{Conclusion}
We presented a general formalism for the classification of superconducting states of matter incorporating time-reversal symmetry. We find it necessary to extend the standard symmetry analysis to keep track of the time-reversal properties by analyzing colored, i.e., Shubnikov,  groups. The specific approach that is conducive to analyze the pairing instabilities is the Shubnikov group of the second kind that  keeps the \change{relative time reversal}  as an explicit symmetry element. We thus  develop an approach in terms of Shubnikov groups of the second kind that allows us to identify odd- and even-frequency solutions within the Bethe-Salpeter equation. In doing so we extend the previous ground-laying work by Volovik and Gor'kov \cite{volovik1984unusual,volovik1985superconducting}, Sigrist and coworkers \cite{ueda1985p,sigrist1991phenomenological} and Blount \cite{blount1985symmetry}. 
Since the combination of spin interchange, parity, and \change{relative time reversal}  has to be odd for a pair of electrons, the found odd-frequency gap symmetries are either both even under spin and even under parity or odd under spin and odd under parity.Consequently, for an experimental identification of the symmetry of a bulk superconducting gap in simple single-band systems, it is required to measure at least two of the three above mentioned information, e.g., as performed in Ref. \cite{PhysRevLett.110.107005}. Even though signals for the experimental verification of odd-frequency states are often discussed in connection to systems which explicitly break time-reversal symmetry \cite{di2015signature,PhysRevX.5.041021,pal2017spectroscopic}, our paper reveals that odd-frequency solutions do not require a time-reversal breaking potential. Odd-frequency solutions can arise naturally for a time-reversal symmetric interaction as a symmetry-breaking ground state of a many-particle system with time-reversal invariant interactions. Although exhibiting a dynamical order, the phenomenon of odd-frequency superconductivity as such is similar to other symmetry-breaking transitions.

\section*{Acknowledgements}
We acknowledge discussions with Sergey Pershoguba, Christopher Triola, Annica Black-Schaffer, Manfred Sigrist, \change{and Benjamin Wieder}. The work at Los Alamos National Laboratory is supported by the US Department of Energy, Grant No. BES E3B7. Furthermore, the work was supported by the Swedish Research Council Grant No.~638-2013-9243, the Knut and Alice Wallenberg Foundation, and the European Research Council (ERC) under the European Union’s Seventh Framework Program (FP/2207-2013)/ERC Grant No.~DM-321031.

\section*{Appendix A: Irreducible representations of $SO(3)$, $O(3)$ and $O(3)\times\left\{E,\op{T}\right\}$}
The group $SO(3)$ contains all proper three-dimensional representation matrices, i.e., orthogonal matrices with determinant $+1$. The spherical harmonics $Y_m^l$ represent a set of basis functions for all irreducible representations of $SO(3)$. To each angular momentum quantum number $l$ belongs a $d_l=2l+1$-dimensional irreducible representation $D^l$. Representation matrices $\mat{D}^l(g)$ for an element $g\in SO(3)$ can be found from the transformation behavior of the spherical harmonics via
\begin{equation}
 \op{g} Y_m^l = \sum_{m'=-l}^l D^l_{m'm}(g) Y_{m'}^l.
\end{equation}
The $D^l_{m'm}(g)$ are also denoted as Wigner $D$ functions. The group $O(3)$ contains all orthogonal matrices with determinant $\pm1$. Hence, it incorporates all proper and improper rotations, i.e., rotations, reflections and the inversion $I$. The inversion acts on the spherical harmonics as
\begin{equation}
 \op{P}(I) Y_m^l = (-1)^l Y_{m'}^l.
\end{equation}
However, since $O(3)$ can be written as the semi-direct product $O(3) = SO(3)\times\left\{E,I\right\}$, we can construct odd ($u$) and even ($g$) representations with respect to inversion via
\begin{align}
 D^l_g(g) &= D^l_u(g) = D^l(g), \\
 D^l_u(Ig) &= -D^l(Ig), \\
 D^l_g(Ig) &= D^l(Ig),
\end{align}
where $g$ denotes a proper rotation, i.e., $g\in SO(3)\subset O(3)$. The spherical harmonics are basis functions of $D^l_g$ for even values of $l$ and basis functions of $D^l_u$ for odd values of $l$. Incorporating the \change{relative time-reversal $\op{T}$}, a similar strategy can be applied to construct representations of $O(3)\times\left\{E,\op{T}\right\}$. It follows for an element $h\in O(3) \subset O(3)\times\left\{E,\op{T}\right\}$ that
\begin{align}
 D^l_{x,+}(h) &= D^l_{x,-}(h) = D_x^l(h), \\
 D^l_{x,-}(Tg) &= -D_x^l(Tg), \\
 D^l_{x,+}(Tg) &= D_x^l(Tg),
\end{align}
where $x=u,g$. 

\change{
\section*{Appendix B: Transformation behavior under relative time-reversal}
We discuss the transformation behavior under relative time-reversal for the anomalous Green function $F$, given by
\begin{equation}
 F_{\sigma\sigma'}\left(\vec{k},t_1,t_2\right) = \left<\mathcal{T} c_\sigma\left(\vec{k},t_1\right) c_{\sigma'}\left(-\vec{k},t_2\right) \right>.
\end{equation}
Here, the operator $\mathcal{T}$ denotes the time-ordering operator, i.e.,
\begin{multline}
 F_{\sigma\sigma'}\left(\vec{k},t_1,t_2\right) =
\left<\theta(t_1-t_2) c_\sigma\left(\vec{k},t_1\right) c_{\sigma'}\left(-\vec{k},t_2\right) \right. \\- \left. \theta(t_2-t_1) c_{\sigma'}\left(-\vec{k},t_2\right)c_\sigma\left(\vec{k},t_1\right) \right> 
\label{appb:eq1}
\end{multline}
Reversing $t_1$ and $t_2$ leads to 
\begin{multline}
 F_{\sigma\sigma'}\left(\vec{k},t_2,t_1\right) =
\left<\theta(t_2-t_1) c_\sigma\left(\vec{k},t_2\right) c_{\sigma'}\left(-\vec{k},t_1\right) \right. \\- \left. \theta(t_1-t_2) c_{\sigma'}\left(-\vec{k},t_1\right)c_\sigma\left(\vec{k},t_2\right) \right>.
\label{appb:eq2}
\end{multline}
Hence, by comparing \eqref{appb:eq1} and \eqref{appb:eq2}, one obtains 
\begin{equation}
F_{\sigma\sigma'}\left(\vec{k},t_2,t_1\right) = -F_{\sigma'\sigma}\left(-\vec{k},t_1,t_2\right).
\end{equation}
Since the gap $\mat{\Delta}$ is related to $\mat{F}$, a similar transformation behavior is present,
\begin{equation}
\Delta_{\sigma\sigma'}\left(\vec{k},t_2,t_1\right) = -\Delta_{\sigma'\sigma}\left(-\vec{k},t_1,t_2\right).
\label{appb:eq3}
\end{equation}
For a spin singlet, the gap can be written as $\mat{\Delta}^{\text{s}}(\vec{k})=i\Psi(\vec{k})\mat{\sigma}_y$, 
\begin{equation}
 \mat{\Delta}^{\text{s}}(\vec{k}) = \left(
 \begin{array}{cc}
  0 & \Psi(\vec{k}) \\
  -\Psi(\vec{k}) & 0
 \end{array}
 \right).
\end{equation}
Hence, from \eqref{appb:eq3} it follows
\begin{align}
 \op{T}\mat{\Delta}^{\text{s}}(\vec{k})&=i\Psi(-\vec{k})\mat{\sigma}_y \\
 \op{T}&:\Psi(\vec{k})\rightarrow \Psi(-\vec{k}).
\end{align}
As a result, the action of $\op{T}$ is similar to the action of the parity operator $\op{P}$ for a spin singlet gap. The product of both is equal to the identity,
\begin{equation}
 \text{singlet:}\qquad \op{T}\op{P} = \op{1}.
\end{equation}
For a spin triplet, the gap is given by $\mat{\Delta}^{\text{t}}(\vec{k})=i \left(\vec{d}(\vec{k})\cdot \vec{\mat{\sigma}}\right)\cdot\mat{\sigma}_y$, 
\begin{equation}
 \mat{\Delta}^{\text{t}}(\vec{k}) = \left(
 \begin{array}{cc}
  -d_1(\vec{k}) + i d_2(\vec{k}) & d_3(\vec{k}), \\
  d_3(\vec{k}) & d_1(\vec{k}) + i d_2(\vec{k})
 \end{array}
 \right).
\end{equation}
Applying equation \eqref{appb:eq3} leads to
\begin{align}
 \op{T}\mat{\Delta}^{\text{t}}(\vec{k})&=-i \left(\vec{d}(-\vec{k})\cdot \vec{\mat{\sigma}}\right)\cdot\mat{\sigma}_y, \\
 \op{T}&:\vec{d}(\vec{k})\rightarrow -\vec{d}(-\vec{k}).
\end{align}
Hence, for the spin triplet, the action of $\op{T}$ differs from the action of $\op{P}$ by a minus sign, and the product of both operators is equal to minus the identity,
\begin{equation}
 \text{triplet:}\qquad \op{T}\op{P} = -\op{1}.
\end{equation}
As the action of $\op{T}$ can be mediated entirely in spin and $\vec{k}$-space, a similar transformation behavior is revealed for reversing $\omega$ after a Fourier transform of the relative time.
}
\bibliography{references}

\end{document}